\documentclass[11pt]{article}
\usepackage{moriondpage,epsfig}
\def\be{\begin{equation}}
\def\ee{\end{equation}}
\def\bea{\begin{eqnarray}}
\def\eea{\end{eqnarray}}

\begin{document}
\title{HEAVY FLAVORS RESULTS FROM SLD\footnote{Invited talk presented at the XXXVIIth Rencontres de Moriond, Les Arcs, March 9-16, 2002.}}

\author{ AARON S. CHOU for the SLD Collaboration\footnote{Work supported in part by Department of Energy contract DE-AC03-76SF00515.}}

\address{Stanford Linear Accelerator Center, 2575 Sand Hill Road,\\
Menlo Park, CA, USA}

\maketitle

\abstracts{We present recent measurements by SLD of the
branching fractions of $B$ hadrons to states with 0 and 2 open charm hadrons,
$BR_{0D}$ and $BR_{2D}$, from which both the average charm yield per $B$ decay,
$N_c$ and the inclusive branching ratio into rare modes not containing
any charmed hadrons, $BR_{rare}$ can be derived.  We also present a
new measurement of the $B_d$ mixing frequency $\Delta m_d$ and limits on
the $B_s$ mixing frequency $\Delta m_s$.  These analyses take
advantage of the excellent vertexing resolution of the VXD3, a pixel-based
CCD vertex detector, which enables the topological separation of the B and
cascade D decay vertices.}

\section{$B$ Decay Charm Counting}

This measurement is motivated by the possible discrepancy between
measurements and Standard Model predictions of the $B$ semileptonic
branching fraction and the average charm quark/antiquark yield per
$B$ decay, $N_c$.  Recent next-to-leading-order calculations
\cite{Neubert:1997we} agree with experiment only at
rather low renormalization scales where it is not obvious that
the perturbation series is nearing convergence.  See 
Yamamoto\cite{Yamamoto:1999va}
for a discussion.  A true discrepancy between the measurements and
the Standard Model predictions could signal either a breakdown of
standard calculational techniques, or an enhancement of the rate of decays
into rare or unexpected new modes.

In this paper, we present measurements using a novel vertexing technique,
of the branching ratios of $B$ hadron decays into final states with
0 or 2 weakly decaying charmed hadrons.  
These decay categories will be referred to as $0D$ decays and $2D$
decays in which $D$ refers to any of $D^0/D^\pm/D^0_s/c$-baryon.
The sample of $B$ hadrons used is the $B^\pm/B^0/B^0_s/b-$baryon admixture
produced in decays of the $Z^0$.  
The measurements of $BR_{0D}$ and $BR_{2D}$ may be combined with
previous measurements\cite{ref:hfcombined} of the $B$ decay branching
ratio into final states
including charmonium, $BR_{c\bar c}$ to get a experimental value for
$N_c$ using:
\begin{eqnarray} 
N_c &=& 1 \times BR_{1D} + 2 \times BR_{2D} + 2 \times BR_{c\bar c} \nonumber \\
    &=& 1.0 - BR_{0D} + BR_{2D} + 2 \times BR_{c\bar c}. 
\label{E:ncj}
\end{eqnarray}  
The second equation here is obtained by using
$BR_{0D} + BR_{1D} + BR_{2D} \equiv 1.0$.  The branching ratio into
rare final states not containing any charm hadrons may also be calculated
as:
\begin{equation}
BR_{rare} = BR_{0D} - BR_{c\bar c}
\label{E:rare}
\end{equation}
These measurements therefore not only yield the average charm count
$N_c$, but also elucidate the composition of the non-semileptonic
decay width.  In addition, the systematic uncertainties in these measurements
are mostly uncorrelated with those of most previous measurements which
have relied on counting $D$'s via reconstruction of exclusive decay modes.
These measurements therefore provide important new data on inclusive
properties of $B$ decays.  

\subsection{The Method}
The measurements of $BR_{0D}$ and $BR_{2D}$ described here utilize
a correspondence between the number of heavy hadron weak decays in an event and
the number of distinct topological decay vertices that can be reconstructed
in the detector.  A $0D$ decay should produce a single
secondary $B$ decay vertex in addition to the primary $Z^0$ decay vertex.
A $1D$ decay should produce an additional secondary vertex at the $D$ decay
position, and a $2D$ decay should produce two additional secondary vertices. 

The correspondence between the number of heavy hadron weak decays and the
number of reconstructed vertices is not perfect due to vertex
finding inefficiencies due to low vertex track multiplicities and finite
tracking resolution.  Tails in the tracking resolution distribution may also
cause false vertices to be formed away from the true decay positions,
resulting in vertex finding overefficiencies.
Due to these difficulties, it is not possible to identify the topological
category of decays on an event-by-event basis.  However, a counting
analysis is still possible, using a suitable unfolding procedure to 
account on average for the effects of these vertexing issues.

The $nD$ branching ratios may be simultaneously measured by simply
counting the number of vertices found in each $B$ decay, and fitting the
secondary vertex count ($N_{vtx}$) distribution for the entire data set
to a linear combination of a set of distributions predicted by the
MC for each decay category.  In order for this procedure to work,
both the MC detector and physics simulations must be carefully tuned to 
accurately model the vertex finding efficiency so that the 
correct probability distribution functions for $N_{vtx}$
for each decay category are produced.  The MC detector simulation is
calibrated using a number of supplementary measurements.
The MC physics simulation is tuned to measurements made by MARKIII, CLEO,
and LEP experiments.  

\subsection{Fitting the vertexing distributions}
To form the vertex count distributions, first a sample of hadronic
$Z^0$ decays is selected by requiring at least 7 measured tracks in 
the drift chamber and a visible energy of at least 30 GeV in the 
calorimeters.  The interaction point (IP) where the $Z^0$ decays 
is measured using the reconstructed tracks.
$Z^0\rightarrow b \bar b$ events are then selected by requiring at least
one $B$-tagged\cite{Wright:2000kv} hemisphere in each event.  This tag
yields a $98\%$ pure sample of $b \bar b$ events.  To get a sample
of generic $B$ decays, only the decays in the hemisphere opposite a
$B$-tagged hemisphere are chosen.  In the resulting sample of $B$ decay
hemispheres, the ZVTOP
\cite{Abe:bsghost} ghost track algorithm is used to reconstruct the $B$ and
cascade $D$ vertices.  

\begin{figure}[htbp]
 \begin{minipage}[t]{0.47\textwidth}
 \resizebox{\textwidth}{!}{\includegraphics{nsvnorm.epsi}}
 \caption{\label{F:mcvtxcnt}MC predicted found secondary vertex count.  Each histogram has been normalized to unit weight.  }
 \end{minipage}
 \hfill
 \begin{minipage}[t]{0.47\textwidth}
 \resizebox{\textwidth}{!}{\includegraphics{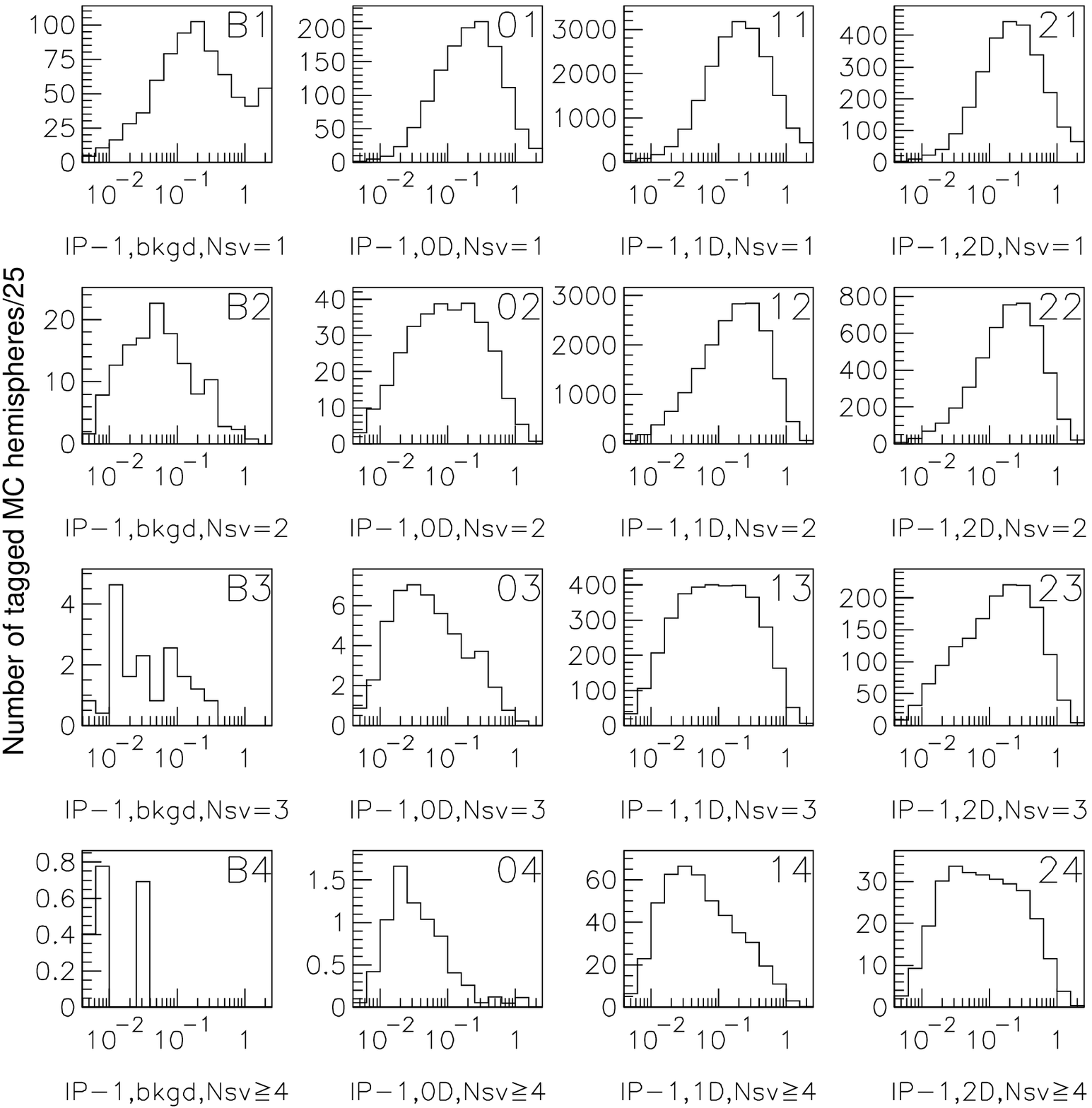}}
 \caption{MC predicted IP-vertex 1 separations (cm) on a log scale.  The rows represent $N_{sv}$=1,2,3,$\ge$4, and the columns represent $udsc$ background, $0D$, $1D$, and $2D$.  The histograms have been normalized to a data-sized sample.}
 \label{F:ip1log}
 \end{minipage}
 \end{figure}

The vertex count distribution measured in the data is fit to a linear
combination of distribution shapes predicted by the MC for each of the
$udsc$ background in the $B$-tagged sample, the $0D$ decays,
the $1D$ decays, and the
$2D$ decays.  The vertex count distribution shapes for each of these
categories are shown in figure~\ref{F:mcvtxcnt}.
These shapes show significant
differences which provide high analyzing power for the fit.  The $0D$
shape is strongly peaked in the 1-vertex bin as expected.  Both a $B$
and a $D$ vertex are found about $50\%$ of the time in the $1D$ category,
and a second $D$ is also found a good fraction of the time in the $2D$
category.  In this last case, due to low vertex track multiplicities,
even an infinitely precise detector could find all three vertices 
at most $\sim 40\%$ of the time. 

In order to utilize
extra available information from the measured vertex positions, these
vertex count distributions are expanded as follows.
The measured $B$ decay length, defined as
the distance between the measured IP position and the nearest reconstructed
vertex in the hemisphere, is histogrammed 
separately for each value of the secondary vertex count.  These histograms,
shown in figure~\ref{F:ip1log},
can be viewed as slices of a 2-dimensional
histogram with $N_{vtx}$ on one axis and the measured $B$ decay length
on the other axis.
These new shape distributions still provide information on the number of $D$'s
through $N_{vtx}$ and but now lifetime information from the long distance
behavior of the decay length distributions as well as vertexing resolution
information from the short distance behavior of the decay length
distributions is also included.  

As before, the 2-dimensional vertex
count distribution is fit to a linear combination of the 
2-dimensional distribution shapes predicted by the MC for each of the four
decay categories.  The fit is a binned $\chi^2$ fit.  Each of the
$N_{vtx}=$1, 2, 3,$\ge 4$ decay length distributions is divided into
14 bins.  In order to provide a balance between the information from
short distance scales and that from long distance scales, the bins are
chosen to be uniform in the logarithm of decay length.  One
extra bin is used for the $N_{vtx}=0$ count where there is no explicit
decay length measurement. The fitting function used is:
\begin{equation}
F_{data}^{i} = R_{n} \cdot [(1-R_{bkgd})\cdot[BR_{0D} \cdot F_{0D}^{i}
 + (1-BR_{0D}-BR_{2D}) \cdot F_{1D}^{i} + BR_{2D} \cdot F_{2D}^{i}]
 + R_{bkgd} \cdot F_{bkgd}^{i}]
\end{equation}
where $F_{0D}, F_{1D}, F_{2D}, F_{bkgd}$ are the four normalized
MC distributions, and i = \{1..57\} is the bin number.  The parameters
extracted from the fit are
normalization $R_{n}$, the $udsc$ background fraction $R_{bkgd}$ in
the $B$ tag and
the branching ratios $BR_{0D}, BR_{2D}$.  $BR_{1D}$ has been eliminated to
impose the constraint that the branching fractions sum to unity.

The result of the fit, including the contributions to the measured
shapes from the various sources, is shown in figure~\ref{F:fitlog}.
The measured distribution appears to be modelled fairly well although
the fit $\chi^2$/d.o.f. = 1.6 is rather large.  Since the detector
resolution has been calibrated in several ways, the remaining
discrepancies are  believed to be due to imprecise modelling of the momentum
spectrum of daughter particles at each decay stage.  Variations of
the MC modelling of the $B$ decays are included in the systematic errors.

\subsection{The Results}
\begin{figure}[htbp]
 \begin{minipage}[t]{0.47\textwidth}
 \resizebox{\textwidth}{!}{\includegraphics{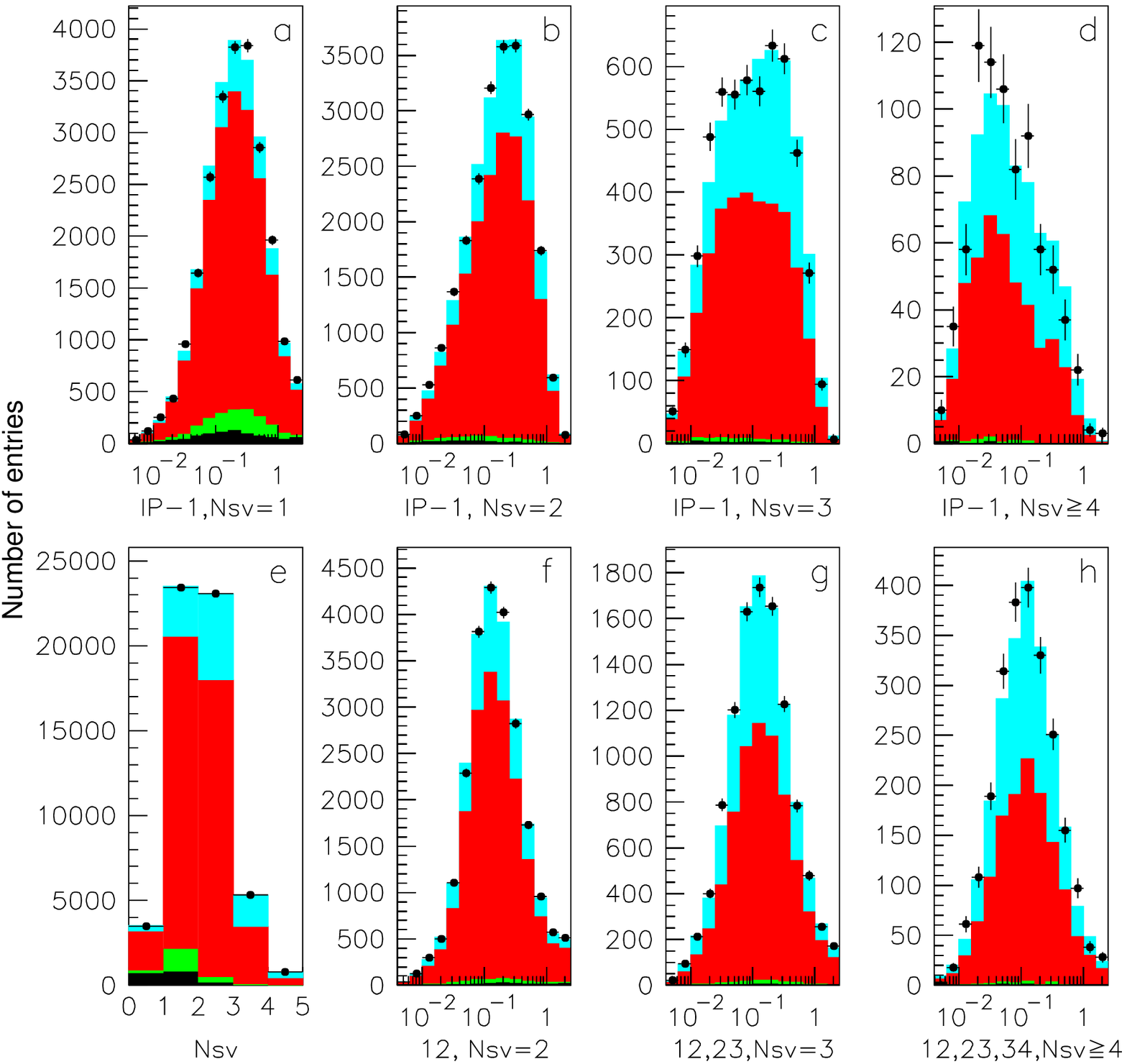}}
 \caption{Fit results on a log scale.  The stacked MC histograms are from bottom to top: $udsc$ background, $0D$, $1D$, and $2D$.  (a)-(d) show the IP to vertex 1 separation [cm],  (f)-(h) shows the nearest neighbor vertex separations [cm], and (e) shows the resulting match in the vertex count distribution. }
 \label{F:fitlog}
 \end{minipage}
 \hfill
 \begin{minipage}[t]{0.47\textwidth}
 \resizebox{\textwidth}{!}{\includegraphics{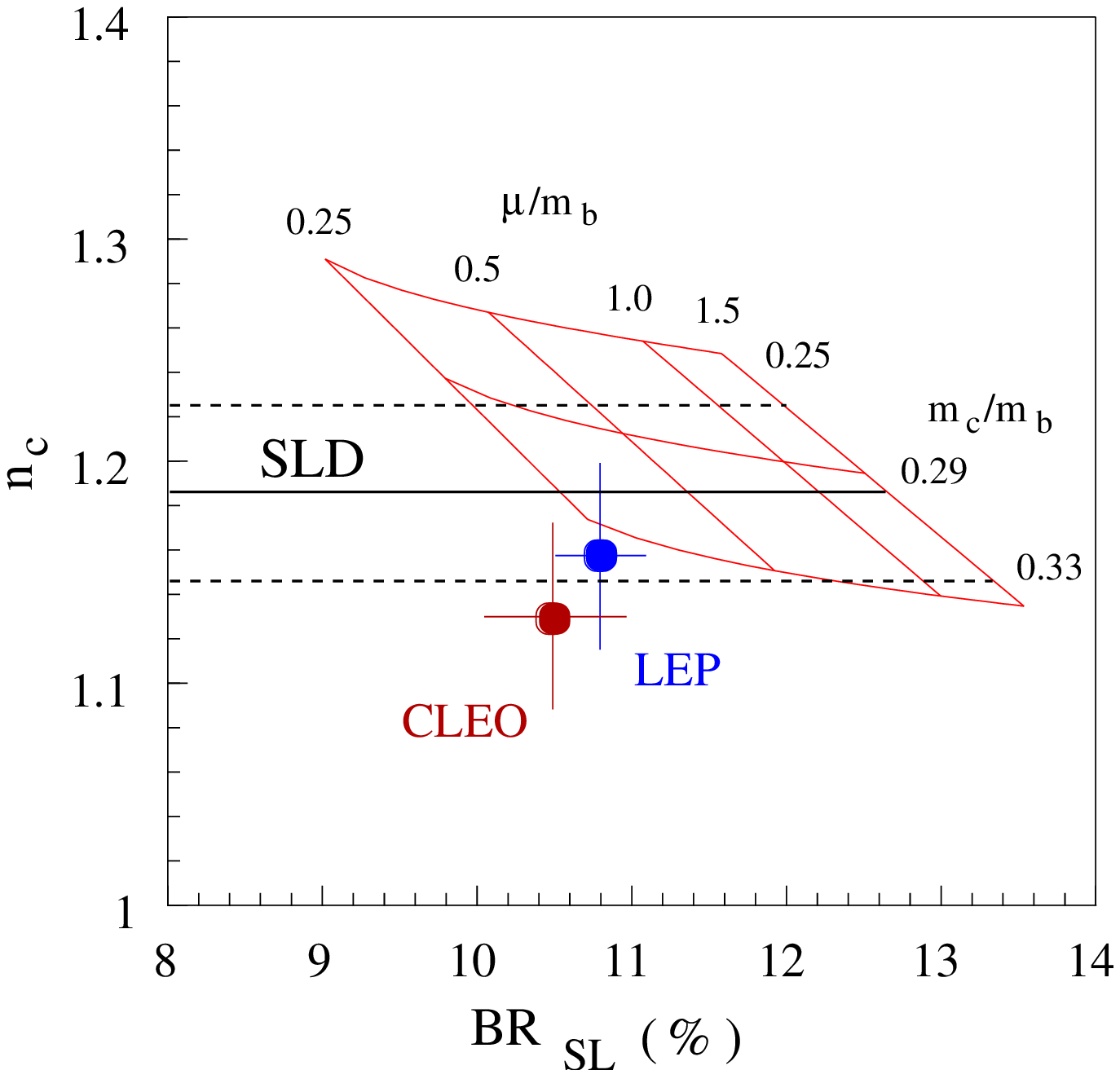}}
 \caption{The SLD measurement of $N_c$ compared with measurements of $N_c$ and $BR_{SL}$ by LEP and CLEO and with theoretical expectations.  The new SLD $N_c$ result is shown as a horizontal band on this plot.  }
 \label{F:measresults}
 \end{minipage}
 \end{figure}

The results of the measurement are:
\begin{eqnarray}
BR(B \rightarrow (0D) X) &=& ( 3.7 \pm 1.1 \pm 2.1)\% \\
BR(B \rightarrow (2D) X) &=& (17.9 \pm 1.4 \pm 3.3)\%
\end{eqnarray}
where the first error is statistical and the second is 
systematic\cite{Chou:2001gj}.  The correlation coefficients between the two
measurements are
$C_{0D,2D} = 0.702$ and $-0.080$ for statistical and systematic errors,
respectively.
$N_c$ is calculated using a value\cite{ref:hfcombined} of
$BR_{(c \bar c)}=(2.3\pm 0.3)\%$ in equation~\ref{E:ncj}:
\begin{equation}
N_c = 1.188 \pm 0.010 \pm 0.040 \pm 0.006.
\end{equation}
Here, the third error is due to the uncertainty in $BR_{c \bar c}$.
The measured value of $N_c$ is plotted in figure~\ref{F:measresults} and
compared with the LEP and CLEO measurement averages\cite{Yamamoto:1999va}
(updated with the new estimates for $\Sigma_c^{0,+}$
and charmonium production\cite{ref:hfcombined}),
and with the theory predicted region\cite{Neubert:1997we}.
The plot indicates that the region of consistency is still at a low
renormalization scale $\mu$.

Limits on $BR_{rare}$ may be set using equation \ref{E:rare}, yielding
a value of $BR_{rare} = (1.4 \pm 2.4)\%$ consistent with the theoretical
expectation\cite{Buchalla:1995kh} of $(2.6 \pm 1.1)\%$ .

\section{$B$ Mixing}
The magnitude of the CKM matrix element $V_{td}$ may be extracted from
measurements of the mixing frequencies $\Delta m_d$ and $\Delta m_s$.
The ingredients of the mixing analyses include
1) a selection of neutral $B$ decays from the $B$-tagged sample described
above; 2) an initial state flavor tag and 3) a final state flavor tag in order
to determine whether a $B$ has mixed before decaying; and 4) a decay 
length and boost measurement to yield the proper time of the decay.  
For $\Delta m_s$, the mixing frequencies are then measured by performing
amplitude fits on the plots of the mixed fraction as a function of proper
time to extract the amplitude of each Fourier mode.  A likelihood fit to
the mixed fraction plot is used to extract $\Delta m_d$.  Here we
report the results of four analyses, three for $\Delta m_s$ and one for
$\Delta m_d$.  More information may be found elsewhere\cite{Abe:bsghost}
\cite{Abe:2000kc} \cite{Wittlin:2001fw}.

All four analyses share the same techniques for obtaining the neutral $B$
sample, for doing the initial state tag, and for measuring the
$B$ boost.  The charge of the decaying $B$ is measured by associating 
charged tracks with a reconstructed $B$ decay vertex to compute the
net vertex charge.  Selecting the zero charge bin yields a sample which
is $87\%$ true neutral decays. 

The initial state flavor tag is naturally provided by forward-backward
asymmetry in the $Z^0$ decay due to the $73\%$ polarization of the
SLC $e^-$ beam.  Other opposite hemisphere information such as the vertex
charge, the momentum weighted jet charge, the charges of identified kaons
and/or high $P_t$ leptons, and the dipole charge (see below), 
are also used in a neural network optimized tag to achieve a mistag
rate of only $22\%$.

The $B$ boost is measured using a combination of calorimetry and of the
kinematics of the reconstructed $B$ vertex.  
$\sigma_p/p \approx 0.08$ for the $60\%$ Gaussian core resolution, and
$\approx 0.20$ for the tail resolution.  

The techniques for the final state tag and decay length 
measurement are described below.  For $\Delta m_s$
the analyses are prioritized in the order below from highest to
lowest sensitivity.  Each $B$ decay is assigned to the highest
sensitivty analysis that it qualifies for. 

\begin{table}[t]
\caption{Comparison of $\Delta m_s$ analyses in the number of events
assigned to each analysis, the core and tail decay length resolution,
the $B_s$ fraction in each event sample, and the final state mistag rate
$w$.\label{T:bsmix}}
\vspace{0.4cm}
\begin{center}
\begin{tabular}{|l|l|l|l|l|}
\hline

& events & $\sigma_L (\mu{\rm m})$ & $B_s$ fraction & w \\ 
& & core (tail) & & \\ 
\hline
Charge Dipole & 11462 & 81 (297) & $16\%$ & $22\%$ \\
\hline
Lepton + $D$  &  2087 & 54 (213) & $16\%$ &  $4\%$ \\
\hline
$D_s^\pm$ + Tracks & 361 & 50 (151) & $38\%$ & $10\%$ \\
\hline
\end{tabular}
\end{center}
\end{table}

\subsection{The $\Delta m_s$ Analyses}
This `$D_s^\pm$ + Tracks' method utilizes a fully 
reconstructed $D_s^\pm$ decay to identify
the sign of the $b$ quark.  
Requiring a $D_s^\pm$ in the event enhances
the $B_s$ fraction by rejecting many of the $B_d$ decays.  The
reconstructed $D_s^\pm$ trajectory
is intersected with the remaining tracks from the $B$ decay to get
a precise decay length measurement.  Because of resulting precise proper
time resolution, this method contributes the most analyzing power at
large $\Delta m_s$ despite having low statistics.  

The `Lepton+$D$' method uses a high-$P_t$ identified lepton from the $B$ decay
to very cleanly identify the sign of the $b$ quark.  Selecting semileptonic
decays rejects the decays with a wrong-sign $D_s^\pm$ which dilute the tag
purity of the other two analyses.  The lepton
trajectory is intersected with the inferred trajectory of a reconstructed
$D$ vertex in the event to measure the $B$ decay length.

The most inclusive analysis uses the `Charge Dipole' technique which
exploits the naturally occurring charge dipole in the
$B_s \rightarrow D_s^- X^+$ decay cascade.  In hemispheres with both
the $B$ and the $D$ vertex reconstructed, the charge dipole may be defined
as the charge difference ($Q_D-Q_B$) between the two vertices multiplied
by the vertex separation distance.  $B_s$ ($\bar B_s$) decays will tend to
have negative (positive) values of this quantity.
The proper time is then measured from the decay length, defined
as the distance between the IP and the closest secondary vertex. 

The results of these three analyses are combined to form the
amplitude fit plot shown in figure~\ref{F:ampfit}.  Based on this plot,
a lower bound of $\Delta m_s < 11.1 {\rm ps}^{-1}$ at $95\%$ confidence
level is calculated.  This bound gives an upper bound on 
$|V_{td}|$ in the $\rho-\eta$ plane of the CKM matrix.

\subsection{The $\Delta m_d$ Analysis}
\begin{figure}[htb]
 \centering
 \begin{minipage}[t]{0.40\textwidth}
 \resizebox{\textwidth}{!}{\includegraphics{datamixfit.epsi}}
 \caption{$\Delta m_d$ likelihood fit to the mixed fraction vs. proper time.}
 \label{F:bdmix}	
 \end{minipage}
 \hfill	
 \begin{minipage}[t]{0.47\textwidth}
 \resizebox{\textwidth}{!}{\includegraphics{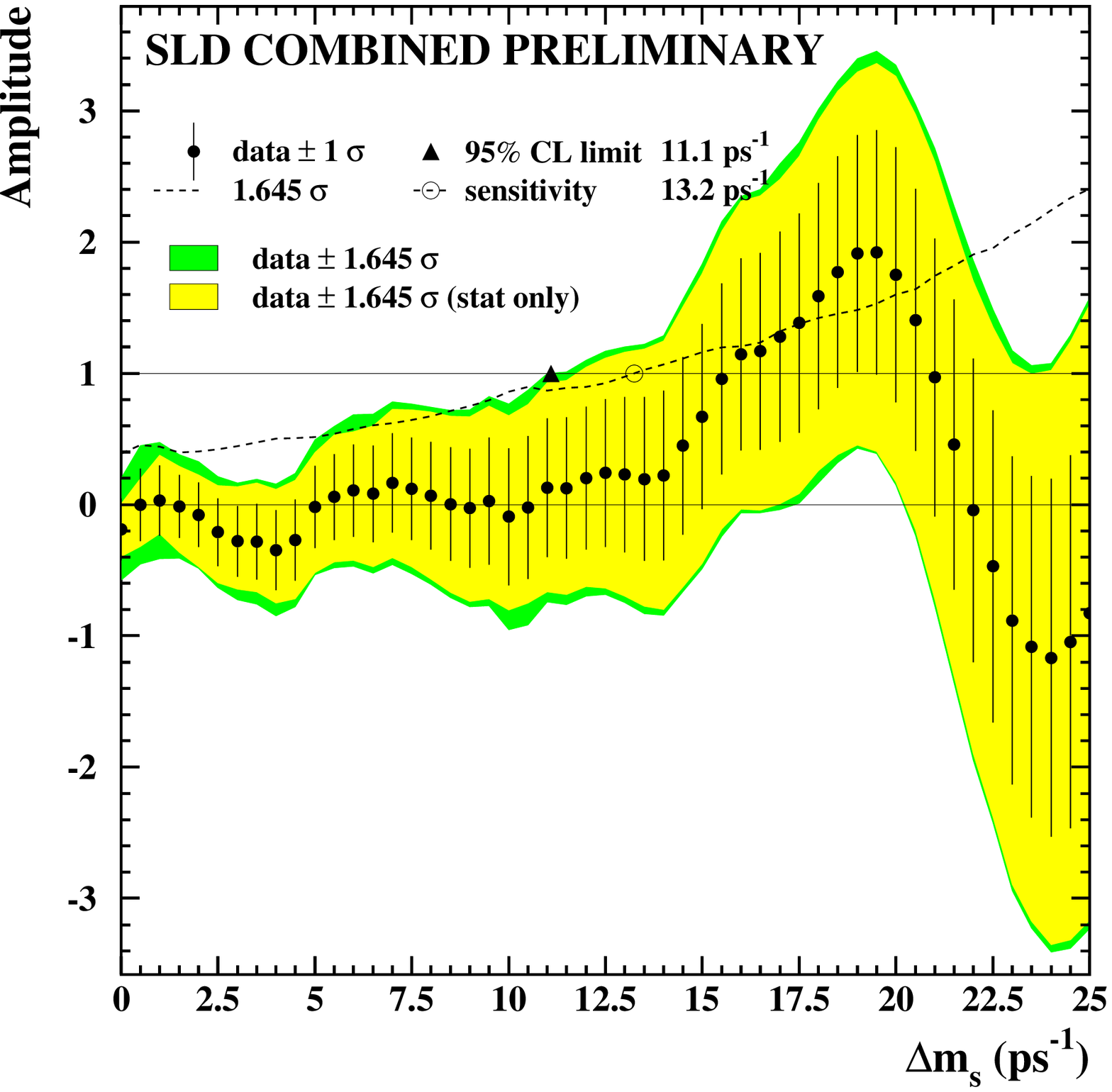}}
 \caption{Combined amplitude fit results for $\Delta m_s$.}  
 \label{F:ampfit}
 \end{minipage}	
 \end{figure}

This analysis tags the sign of the $b$ quark using the identified $K^\pm$
from the cascade $D$ decay.  This tag is calibrated with the data by
simultaneously fitting for $\Delta m_d$ and the `right sign fraction'
in the $B$ decay sample (figure~\ref{F:bdmix}).  The latter fraction
is measured to be $(79.7\pm 2.2)\%$ of the sample of 7844 $B$ decays used.    
The mixing frequency is measured to be 
$\Delta m_d = 0.503 \pm 0.028 \pm 0.020$.

\section*{References}

\bibliography{moriondnotitle}

\end{document}